\documentclass[final,5p,times,twocolumn]{elsarticle}
\usepackage{amssymb}

\journal{Physics Letters B}

\begin{document}

\begin{frontmatter}

\title{Event-by-event charged-neutral fluctuations in Pb+Pb  collisions at 158 A~GeV \\
\large(WA98 Collaboration)} 
%(WA98 Collaboration)} 

%\large{WA98 Collaboration}
\author[PU]{M.M.~Aggarwal} 
\author[VECC]{Z.~Ahammed}
\author[UG]{A.L.S.~Angelis\fnref{fn1}}
%\author[UG]{A.L.S.~Angelis}
\author[KUR]{V.~Antonenko}
\author[JINR]{V.~Arefiev}
\author[JINR]{V.~Astakhov}
\author[JINR]{V.~Avdeitchikov}
\author[ON]{T.C.~Awes}
\author[JAM]{P.V.K.S.~Baba}
\author[JAM]{S.K.~Badyal}
% ...
%\fntext[fn1]{Deceased}
\address[PU]{University of Panjab, Chandigarh 160014, India} 
\address[VECC]{Variable Energy Cyclotron Centre,  Calcutta 700 064, India} 
\address[UG]{University of Geneva, CH-1211 Geneva 4,Switzerland} 
\address[KUR] {RRC Kurchatov Institute, RU-123182 Moscow, Russia} 
\address[JINR]{ Joint Institute for Nuclear Research, RU-141980 Dubna, Russia}
\address[ON]{Oak Ridge National Laboratory, Oak Ridge, Tennessee
  37831-6372, USA} 
\address[JAM]{University of Jammu, Jammu 180001, India} 
% ...
% \vspace*{1cm}

\date{\today}
\begin{abstract}

Charged particles and photons have been measured in central Pb+Pb collisions
at 158~A~GeV in a common 
($\eta$ - $\phi$)-phase space region in the WA98 experiment at the CERN SPS. 
The measured distributions have been analyzed 
%using the sliding window method in order 
to quantify the frequency with which phase space regions of varying sizes 
have either small or large neutral pion 
fraction.
The measured results are compared with VENUS model simulated events  and with 
mixed events. Events with both large and small charged-neutral fluctuations are observed to
occur more frequently than expected statistically, as deduced from mixed events, or as predicted by 
model simulations, with the difference becoming more prominent with decreasing size of the $\Delta\eta$ - $\Delta\phi$ region. \\ 
%\vspace{0.3cm}
 
\end{abstract}

%\PACS{{25.75.-q,}{25.75.Gz}  }% end of PACS codes

\end{frontmatter}

\begin{twocolumn}
%  Key words: The WA98 experiment, photon multiplicity Detector, 
% Disoriented Chiral Condensate,silicon pad multiplicity detector,sliding 
% window method, charged-neutral fluctuations.

Enhanced fluctuation in the production of neutral versus charged pions 
has been one of the predicted signals 
for chiral symmetry restoration in heavy ion collisions at ultra-relativistic
energies~\cite{MOH05}. It has been proposed 
that the extreme energy density
of the matter produced in the region of spatial overlap between two heavy ions 
colliding at relativistic energies may provide the physical 
conditions necessary for the formation of a
chiral condensate that may be aligned in a direction different from the true vacuum.
Domains of such Disoriented Chiral Condensates (DCCs) are expected to emit
pions coherently from the collision volume, which may result in large 
fluctuations in the neutral pion fraction, $f$, defined as $f$ =
%$\frac{N_{\pi^o}}{N_{\pi^o} + N_{\pi^{\pm}}}$,
$N_{\pi^o}/(N_{\pi^o} + N_{\pi^{\pm}})$,
where $N_{\pi^o}$ 
and $N_{\pi^{\pm}}$ are the multiplicities
of the neutral and charged pions, respectively. 
The neutral pion fraction $f$ for DCC domains is predicted to follow a 
probability distribution of the form  
%P($f$) = $\frac{1}{2\sqrt{f}}$, 
P($f$) = $1/2\sqrt{f}$~\cite{blaizot},
which is very different from the binomial (or negative binomial) distribution observed for
generic pion production in hadron collisions. 

Events with large charged-neutral fluctuations, the so called Centauro ($N_{ch}$ $>>$ $N_{\gamma}$)
and Anti-Centauro ($N_{\gamma}$ $>>$  $N_{ch}$)
events, were first observed in the JACEE cosmic ray experiment~\cite{JACEE}. 
The search for such  unusual events and DCC formation 
was later carried out by the D0~\cite{D0}, CDF~\cite{CDF}, and Minimax~\cite{MNMX} 
experiments in p+p
collisions at the Tevatron and by the NA49 ~\cite{NA49} and WA98~\cite{AGG98,AGG01,AGG03} experiments in nuclear 
collisions at the CERN SPS. Upper limits to the 
production of DCC domains have been set within various model assumptions.
WA98 has conducted an exhaustive search for charged-neutral fluctuations in nuclear collisions.
%has been carried out within the WA98 experiment at the CERN SPS.
%using data from the silicon pad multiplicity detector (SPMD)~\cite{SPMD} and the photon multiplicity detector (PMD)~\cite{PMD}. 
In the first WA98 analysis~\cite{AGG98},  
correlated neutral {\it vs.} charged particle fluctuations were investigated globally 
over a  large $\eta$ - $\phi$ acceptance of more than one unit of rapidity near 
mid-rapidity, with full azimuthal coverage. 
Subsequent investigations based on statistical correlations~\cite{AGG01} or
multi-resolution Discrete Wavelet Transform (DWT) techniques ~\cite{AGG03}, searched for DCC-like 
fluctuations in localized regions of ($\eta$ - $\phi$) phase 
space.
% by dividing the azimuthal space in different bins ranging from 2 to 16 
%and upper limits on DCC formation were extracted within a simple model. Such 
%analysis, using fixed bin size in azimuth, have certain limitations of 
%sensitivity in picking up the signals. DCC, if it is formed, may have 
%different domain sizes and may show up at different regions of phase space 
%in different events so that any fixed azimuthal binning would most 
%likely miss it.

In this article we present an improved search for localized 
charged-neutral fluctuations in central Pb + Pb collisions at the SPS using a 
simple  Sliding Window Method (SWM)~\cite{AGG06}. This method utilizes
the full azimuthal resolution of the WA98 detectors to identify all regions 
with unusually large or small values of the neutral particle fraction. 
The sensitivity of the SWM has been studied using simulated data
based on simple model assumptions of charged-neutral fluctuations  
similar to those used in the study of the DWT method~\cite{AGG01,NAN99}. The SWM is found to
 have a sensitivity which is limited only by the available statistics and can
 be several orders of magnitude better than the DWT method for a given data sample. 
Preliminary results have been presented in Refs.~\cite{SOO02,AGG2003}.

The analysis reported here used the measurement of 
photons and charged particles produced in Pb+Pb collisions at 
158 A~GeV recorded by the WA98 experiment, during a period of magnetic field-off 
operation of WA98 in the 1996 run period of the CERN SPS.
Photons were measured with the Photon Multiplicity Detector 
(PMD)~\cite{AGG1999} and 
charged particles were measured with the Silicon Pad Multiplicity Detector (SPMD)~\cite{SPMD}.
 The analysis is  restricted to their common region of $\eta$ - $\phi$ acceptance overlap.
The PMD was located at 21.5~meters downstream of the 
target and consisted of plastic scintillator pads of varying sizes arranged
inside 22 box modules placed behind 3X$_{0}$ thick lead converter plates~\cite{AGG1999}. It covered the pseudorapidity region 2.9~$< \eta <$~4.2. For the photon identification criteria applied
the average photon counting efficiency and the purity of the photon sample were found to be 68\% 
and 65\%, respectively, for the central event sample used~\cite{AGG1999}. The azimuthal resolution 
of the PMD is much less than $1^{\circ}$ and the limit for detection of low $p_{T}$ particles is 30 MeV/c. The SPMD was located at 32.8 cm from the target and
consisted of 22 radial and 46 azimuthal segments in each of four quadrants. It covered the pseudorapidity region of 2.35~$< \eta <$~3.75~\cite{SPMD}. The charged particle detection efficiency of the SPMD was 99\%. 
The azimuthal resolution of 
the SPMD is $2^{\circ}$  and the limit for detection of low $p_{T}$  particles 
is nearly zero. The total transverse energy $E_{T}$ measured in the pseudorapidity region
3.5~$< \eta <$~5.5 by the Midrapidity 
Calorimeter (MIRAC)~\cite{MIRAC},  located 24~m downstream of the target, was used to characterize
the event centrality. All procedures for event selection and removal of backgrounds 
described in detail in previous 
publications~\cite{AGG01,AGG03,SOO02,AGG1999} were followed 
in the present analysis. A total of 185~K events, belonging to the 15\% most central  collisions of
the WA98 minimum bias cross section, have been analyzed.

In the present analysis, the multiplicity of photons measured in the PMD and the multiplicity of
charged particles measured in the SPMD, in the region of common $\eta$ acceptance (2.9~$< \eta  <$~3.75), are employed as experimental observables to approximate the neutral 
pion fraction by the neutral particle fraction 
$f$ $\approx$ $\frac{N_{\gamma}/2}{N_{\gamma}/2 + N_{ch}}$.  
The neutral particle fraction
$f$ is calculated within windows of varying size $\Delta\phi$ in azimuth.

The search for non-statistical fluctuations requires comparison with a model-independent baseline, 
free of dynamical correlations, that includes detector effects and  fluctuations of statistical nature only. 
For the analysis of experimental
data the technique of event-mixing  can provide such a reference data sample in which 
dynamical correlations among particles are completely destroyed, leaving only correlations 
that might result from efficiency variations. The mixed events are 
generated by the standard procedure of using real events to  
construct new mixed events with particles selected randomly, one particle 
per real event. In this analysis, the mixed events are created with the same global multiplicity 
distribution as the experimental data.  Furthermore, for each real event, a mixed event is created with the same global charged particle and photon hit multiplicity, thereby maintaining the same global charged-neutral multiplicity correlation. Two kinds of mixed events are used in this study~\cite{AGG03}. In the first case, termed M1 mixed events, all of the charged particle and neutral hits in each mixed event are selected from different real events. Thus the M1 mixed events constitute fully mixed events with only the global charged-neutral multiplicity correlation maintained. The second class of artificial events, termed M2 mixed events~\cite{AGG03}, are created by mixing PMD and SPMD subevents from different events, 
but without mixing hits within the individual detectors.  
In the M2 mixed events the charged and neutral fluctuations are separately maintained, but the correlation between charged and neutral hits is removed, except that the global charged-neutral multiplicity correlation is maintained. In what
follows the mixed events are the M1 type unless explicitly stated.

Two approaches have been used to characterize the statistical fluctuations and trivial
charged-neutral fluctuations arising in generic particle production. 
Statistical fluctuations due to finite particle effects and small
data samples are studied by a comparison with mixed events generated from 
the data, as mentioned above. 
In addition, the magnitude of trivial charged-neutral fluctuations arising 
in generic particle production have also been
studied with simulated events produced by the VENUS event generator~\cite{VENUS}
passed through the GEANT detector simulation~\cite{GEANT} of the WA98 experiment 
setup, here referred to as VG events, and mixed events generated from these VG
events. For this study, 27k VG simulated Pb+Pb collision events have been generated with impact parameter range selection corresponding to the 15\% most central collisions of the data sample.

\begin{figure}
\centerline{\includegraphics[scale=0.385]{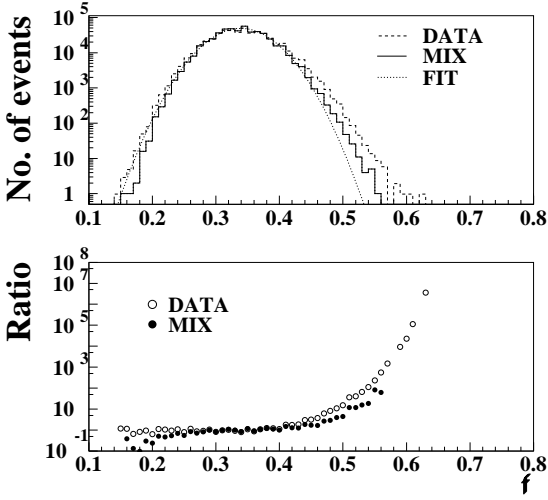}}
\caption{Top: The $f$  distributions for data and mixed events for the 5\% most
central Pb+Pb events for randomly selected $\eta$ - $\phi$ regions with $\Delta\phi$ =60$^\circ$. 
A fit to the distribution for mixed events with a Gaussian function, with fit parameters
 mean = 0.33923$\pm$0.00004 and $\sigma$ =0.04041$\pm$0.00004 , is also shown. Bottom: Ratio of 
the $f$ distributions for the data and the Gaussian fit to the mixed events (open circles) and the ratio of the mixed events with its Gaussian fit (solid circles). }
\label{fig1}
\end{figure}

Fig.~\ref{fig1} (top) shows the $f$ distribution for data for
the 5\% most central Pb+Pb events within different regions of $\Delta\phi$ =60$^\circ$ in an event. 
For comparison, the distribution obtained from M1 mixed events is
also shown together with a Gaussian function fitted to the mixed event distribution.  
It is observed that the $f$ distributions for the
data and mixed events are asymmetric and extend 
beyond the Gaussian fit to larger $f$-values. In Fig.~\ref{fig1} (bottom) 
the ratio of the $f$ distribution for the data to the Gaussian fit to the 
mixed event distribution is shown along with the ratio of the mixed event distribution to its Gaussian fit. The high value of the ratio at large $f$
 signifies vanishing small values of the fitted distribution. 
The $f$ distribution for the data is observed to be broader than that of the mixed events
with small and large $f$-values occurring more frequently.
Because of the predominance of generic pion production 
around $f \approx 1/3$, 
non-statistical fluctuations can best be studied in the tail regions of
the $f$ distribution.

For this reason we focus our attention on the maximum and the minimum $f$-values 
(i.e., $f_{max}$ and $f_{min}$)  amongst all $\Delta\eta$-$\Delta\phi$ regions in each
event in order to search for photon-excess and charge-excess type fluctuations.
This is done using the SWM,
where  the azimuthal plane is scanned by 
sliding the  $\Delta\phi$ window of chosen size in steps of 
$\delta\phi$ = 2$^\circ$ (limited by the azimuthal resolution of the SPMD) and 
computing $f$
for each $\Delta\phi$ region to extract the maximum
and the minimum values of $f$ 
in each event, represented by $f_{max}$ and 
$f_{min}$, respectively. The window size  is kept fixed in $\eta$ and
 varied in azimuth from  
$\Delta\phi=20^\circ$ to 150$^\circ$.
% and the  $f_{max}$ and $f_{min}$ distributions are extracted
%for each $\Delta\phi$ window size.  
The $f_{max}$ and $f_{min}$ distributions for the
experimental data 
and for VG events for the 5\% most central Pb+Pb events  are shown in Fig.~\ref{fig2} for a
window size of $\Delta\phi$ = 60$^\circ$. The distributions for mixed
events generated from data and VG events are also shown. The measured $f_{max}$ and $f_{min}$
distributions are seen to extend further to the right and left, respectively, than the mixed events
distributions, with the difference observed in the data seen to be greater than that for the VG
events.

\begin{figure}
\centerline{\includegraphics[scale=0.385]{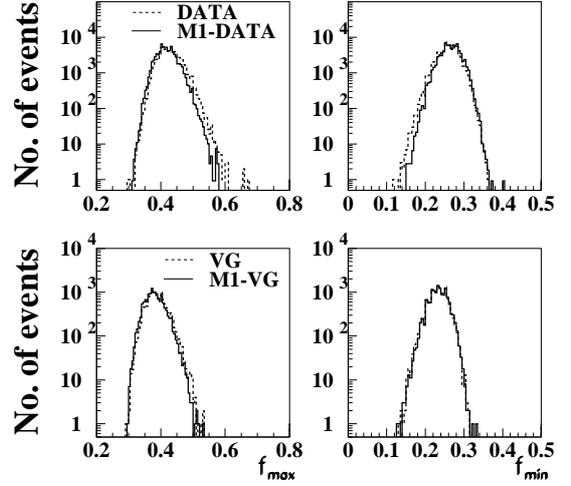}}
\caption{ The $f_{max}$ (left) and $f_{min}$ (right)  distributions for $\Delta\phi$ = 60$^\circ$
window size for data (top) and VG (bottom) for the 5\% most central events together with the distributions for their respective mixed events.}
\label{fig2}
\end{figure}

In order to study the fluctuations in greater detail the
mean ($\mu$) and RMS deviation ($\sigma$) are calculated for the mixed event distributions for both
the experimental data and the VG events. The events in the tails of the $f_{max}$ and $f_{min}$
distributions of the experimental data and the VG events are selected by applying
a cut at  $\mu \pm n\sigma$, with positive sign being applied 
for the $f_{max}$ distribution and negative sign for the $f_{min}$
distribution. Events having regions with $f_{max}$ or $f_{min}$ 
values beyond the cut are labeled as 'exotic' events. Results for  $n$
 values of 3 and 4 are presented.

\begin{figure}
\centerline{\includegraphics[scale=0.385]{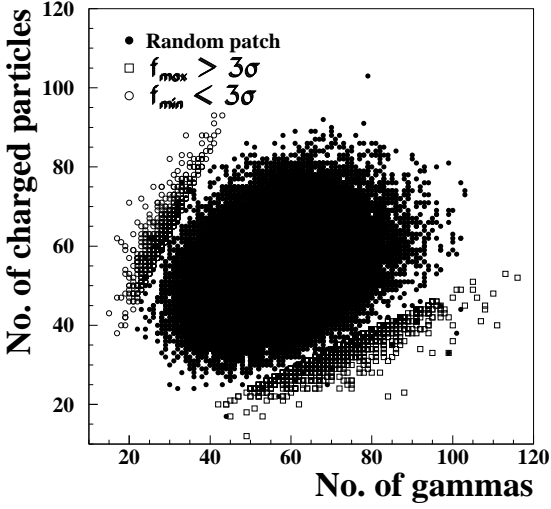}}
\caption{ Scatter plot of N$_{\gamma}$ versus N$_{ch}$ for a randomly
selected  60$^\circ$ region for the 5\% most central Pb+Pb events (solid points).
The N$_{\gamma}$ versus N$_{ch}$ correlation 
for the photon and charge excess regions found by the SWM with the $3\sigma$ cut
for the same event sample is also shown (open points).}
\label{fig3}
\end{figure}

Fig.~\ref{fig3} displays
a scatter plot of N$_{\gamma}$ versus N$_{ch}$ for a randomly selected
$\Delta\phi=60^\circ$ region in each event. A weak correlation of N$_{ch}$
with  N$_{\gamma}$ is observed due to residual impact 
parameter correlations, but with a wide spread. It is observed that a
few points lie beyond the main 
cluster of events corresponding to events with 
large fluctuations in the multiplicities of photons and charged particles. 
When the SWM is applied to the same events to select those events 
where the $f$ values exceed $(\mu + 3 \sigma)$ for $f_{max}$ or 
fall below $(\mu - 3\sigma)$  for $f_{min}$ many more events with 
regions having large fluctuations in the number of
photons and charged particles are found, as shown by the open points in Fig.~\ref{fig3}.
% This is  not possible  by randomly selecting a region i.e., fixed
% azimuthal binning, in $\eta$-$\phi$ region in an event wherein the  
% exotic region splits into different regions in $\eta$-$\phi$ space resulting 
% in diluting the given signal and its significance.

\begin{figure}
\centerline{\includegraphics[scale=0.385]{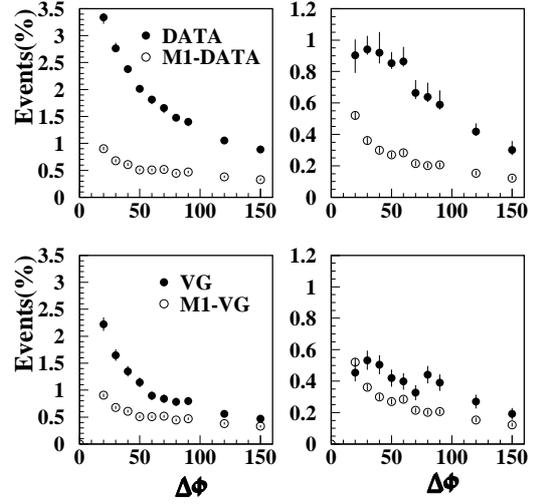}}
\caption{ The percentage of events having $\Delta\eta$-$\Delta\phi$ regions with $f_{max}$ 
exceeding a $3\sigma$ cut (left) and with $f_{min}$ less than a $3\sigma$ cut (right) 
versus the window size $\Delta\phi$ for the 5\%
most central events for data (top) and for VG (bottom) together with the results for their 
respective mixed events. Errors bars denote both statistical and systematic 
contributions. }
\label{fig4}
\end{figure}

Fig.~\ref{fig4} shows the percentage of events  beyond 
the $\mu \pm 3\sigma$ cut as a function of the window size $\Delta\phi$. 
Results are shown for data and mixed events in the top 
panel and for VG events and VG mixed events in the bottom panel. The results 
for the mixed events from data and VG are seen to be very similar. If the $f_{max}$ ($f_{min}$) 
distributions were well described by a Gaussian function, the percentage of mixed events exceeding the 
$3\sigma$ cut is expected to be 0.135\%, independent of $\Delta\phi$. Instead, 
it is observed 
that due to the asymmetry of the $f_{max}$ ($f_{min}$) distribution (see Fig.~~\ref{fig2}) 
there are excess number of events in the 
region of larger (smaller) $f$-values, with an excess that grows with decreasing $\Delta\phi$ bin size, 
even for mixed events.  It is observed
that the percentage of exotic events in the data exceeds that in mixed events
for all  $\Delta\phi$ bin sizes for both the photon-excess and charge-excess selections. 
While the mixed events for data and VG events are very similar, 
the difference between data and mixed events exceeds the difference between 
VG and its mixed events. 
It is also seen that the difference between VG and its mixed events
is much less for the charge-excess case than for the photon-excess case. 
The PMD-SPMD detector combination is better
suited to study the photon-excess fluctuations because the purity of the 
photon sample in the PMD improves for large $f$ and can reach values up to 
85\%~\cite{MOH04}. For the charge-excess case the charged particle contamination plays a
significant role in reducing the purity of the photon sample, which can be as
low as 45\%, and also introduces unwanted correlations due to charged hits in the SPMD 
that are simultaneously mis-identified as photon hits in the PMD.  

The various sources of systematic errors associated with $N_{\gamma}$ and
$N_{ch}$ distributions have been investigated and described in detail
previously~\cite{AGG1999}. The systematic error in the determination of photon
multiplicity is (-7.1\%,+3.4\%) and in the charged multiplicity it is
$\pm$4\%. To investigate the systematic errors on the number of observed
patches new set of event samples was generated by randomly (a) removing 7.1\%
of photon hits and adding 4\% of charged particle hits and (b) adding
3.4\% photon hits and removing 4\% charged particle hits. For these samples
mixed events samples were also generated. These new event
 samples were analyzed
in the same way 
as the real data and mixed
events samples mentioned earlier to obtain the  systematic errors.
 Errors shown in  figures are quadratic
sum of statistical and systematic errors.

\begin{figure}
\centerline{\includegraphics[scale=0.385]{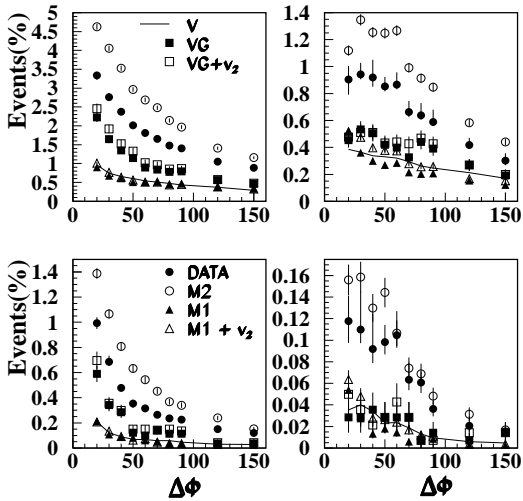}}
\caption{ Percentage of events having regions with $f_{max}$ (left)
exceeding $3\sigma$ (top) or $4\sigma$ (bottom) cuts and  with $f_{min}$ (right)
less than $3\sigma$ (top) or $4\sigma$ (bottom) cuts
versus $\Delta\phi$  for different sample types as described in the text.}
\label{fig5}
\end{figure}

The percentages of events having exotic photon and charge excess regions
have been further investigated for additional types of event samples. 
In order to determine if the larger fluctuations observed in the data might be due to elliptic flow, 
the mixed and VG events have been modified to introduce elliptic flow following the
method of Poskanzer and Voloshin~\cite{POS98}. The p$_{T}$ dependent
values of v$_{2}$ are taken from Ref.~\cite{NA4907}. Since the
p$_{T}$ of the particles is not measured in the PMD or SPMD, a v$_{2}$ value has been assigned
to a given particle 
following the p$_{T}$ probability distribution obtained from the VENUS
event generator. 
Fig.~\ref{fig5} (top) shows the percentage of events in which $f_{max}$ (left)
exceeds the $3\sigma$ cut or in which $f_{min}$ (right) is
less than the  $3\sigma$ cut versus $\Delta\phi$ for M1 mixed events with flow 
(M1+v$_2$) and VENUS+GEANT with flow (VG+v$_2$). 
%The results may be compared with those for the data (DATA), M1 mixed events, and VG events, also shown in Fig.~\ref{fig3}. 
It is seen
that the observed enhancement in the percentage of exotic events in the
data can not be explained as due to elliptic flow. 

The analysis was also performed on M2 mixed events constructed as described above~\cite{AGG01,AGG03} by mixing the unaltered PMD hits of one event with the 
unaltered SPMD hits of a different event, keeping the same total
number of photons and charged particles as in the original event, and hence keeping the
overall correlation between them. 
The results for the M2 event analysis shown in Fig.~\ref{fig5} are seen to lie above the 
data signifying that the charged-neutral fluctuations in the data tend to be correlated and give
rise less frequently to events with photon-excess or charge-excess 
regions than the randomly correlated charged 
and neutral hits of M2 mixed events. This behaviour was previously 
reported in Ref.~\cite{AGG01} and is contrary to the naive expectations for a DCC. The
result might be attributed to residual impact parameter correlations in the charged and neutral
multiplicity in the data.

The curves shown in Fig.~\ref{fig5} were obtained from 
VENUS simulations without GEANT detector response
for the 5\% most central collisions corresponding to the impact parameter range
0-3.5 fm. The pure VENUS events are seen to follow closely the M1 mixed event results indicating
essentially no correlations in the particle production in the VENUS event generator.  The increase in 
the percentage of events with neutral or charge excess after application of 
the detector response (VG) is attributed to correlation due to detector 
effects where some charged particles may be mis-labeled as photons in the PMD.

The bottom panels of Fig.~\ref{fig5} show the same analysis as in the upper panels but with the cuts on the $f_{max}$ (left) and $f_{min}$ (right) distributions increased to $4\sigma$.
Although the percentage of events in the photon-excess or charge-excess regions are 
decreased  with the more stringent cuts, as expected,
the general trends and conclusions are the same when comparing the results for the $3\sigma$ cuts (Fig.~\ref{fig5} (top)) with those for the $4\sigma$ cuts (Fig.~\ref{fig5} (bottom)). 

The analysis has also been performed for events in the 5-10\% and 10-15\% most
central event selections. Table~\ref{t1} compares these results for the $4\sigma$ cut with the 0-5\% results
for different event sample types as discussed above. This comparison shows little  centrality 
dependence for the results over this limited range of centralities,
 there being a slight tendency of 
increased excess in the data with increasing centrality.

\begin{table}
\caption{\label{tab:table} Percentages of exotic events for the $4\sigma$ cut
 on the $f_{max}$ (photon excess)  or $f_{min}$ (charge excess) 
for Data, M1 mixed, Venus + Geant (VG) and M2 mixed
events for different centralities for the $\Delta\phi=60^{\circ}$ window size. The statistical
and systematic errors are indicated.}
\label{t1}
\begin{center}
\vspace{0.3cm}
\begin{tabular}{|l|l|l|r|} \hline

{Cent.} & {Sample}&$f_{max}$&$f_{min}$ \\ \hline
0-5\% &DATA&0.316$\pm$0.020&0.105$\pm$0.012 \\
      &SYS ER&+0.005-0.015&+0.019-0.000 \\
      &M1&0.081$\pm$0.010&0.014$\pm$0.004 \\ 
      &VG&0.100$\pm$0.027&0.028$\pm$0.014 \\
      &M2&0.471$\pm$0.027&0.083$\pm$ 0.011 \\ \hline
5-10\% &DATA&0.260$\pm$0.019&0.055$\pm$0.009 \\
      &SYS ER&+0.016-0.006&+0.011-0.003 \\
      &M1&0.048$\pm$0.008&0.013$\pm$0.004 \\
      &VG&0.099$\pm$0.033&0.000$\pm$0.000 \\
      &M2&0.471$\pm$0.028&0.113$\pm$0.014 \\ \hline
10-15\% &DATA& 0.274$\pm$0.028&0.045$\pm$0.011 \\
       &SYS ER&+0.023-0.020&+0.00-0.014 \\
       &M1&0.051$\pm$0.0140&0.014$\pm$0.007 \\
       &VG& 0.074$\pm$0.033&0.015$\pm$0.015 \\
       &M2&0.357$\pm$0.032&0.083$\pm$0.016 \\ \hline

%\hline
%{Cent.} & {Sample} & $f_{max}$ & $f_{min}$ \\ \hline
%0-5\% &DATA &1.81$\pm$0.05&0.87$\pm$0.03 \\
%      &SYS ER&+0.04-0.07&+0.09-0.02 \\
%      &M1 &0.51$\pm$0.03&0.28$\pm$0.02 \\
%      &VG &0.90$\pm$0.08&0.40$\pm$ 0.05 \\
%      &M2&2.41$\pm$0.06&1.11$\pm$0.04 \\ \hline\hline
%5-10\%&DATA&1.55$\pm$0.05&0.65$\pm$0.03 \\
%      &SYS ER & +0.01-0.05&+0.07-0.01 \\
%      &M1&0.48$\pm$0.03&0.23$\pm$0.02 \\
%      &VG&0.94$\pm$0.10&0.24$\pm$0.05  \\
%      &M2&2.69$\pm$0.07&1.15$\pm$0. 04 \\ \hline\hline
% 10-15\%&DATA&1.48$\pm$0.06&0.59$\pm$0.04 \\
%      &SYS ER &+0.03-0.07&+0.03-0.03 \\
%       &M1&0.50 $\pm$0.04&0.22 $\pm$0.03 \\
%       &VG&0.65$\pm$0.10&0.36$\pm$0.07 \\
%       &M2&2.09 $\pm$0.08&0.89$\pm$0.05 \\ \hline\hline
%
\end{tabular}
\end{center}
\end{table}

In addition to the event selection and data clean-up methods that have been employed,
a set of additional checks have been performed with the 5\% most central
collisions for the $\Delta\phi=60^{\circ}$ window size to further examine the 
event structure and to search for possible detector artifacts.

\begin{figure}
\centerline{\includegraphics[scale=0.39]{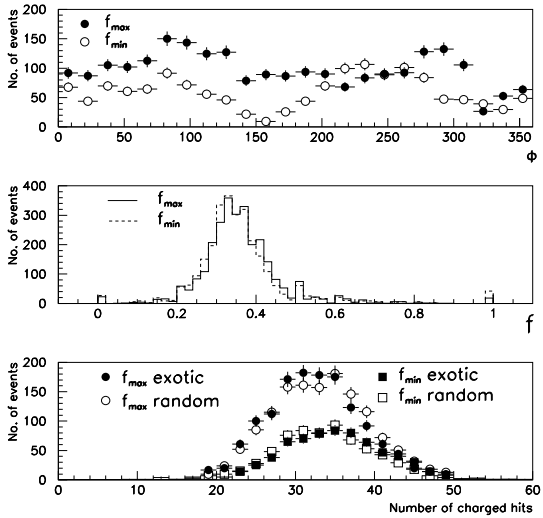}}
\caption{Results of various checks: (top) azimuthal distributions of exotic
photon and charge excess regions; (center) $f$ distributions for  
preceding and succeeding events  for the same region size at the same 
azimuth as indicated; and (bottom) charged particle multiplicity
distribution in the non-overlapping $\eta$ region of the SPMD at the same azimuthal 
position as the exotic region and in a randomly selected region of the SPMD of
similar size in the same event.}
\label{fig6}
\end{figure}

The first check concerns the azimuthal distribution of the exotic regions. 
Fig.~\ref{fig6} (top) shows the azimuthal distribution 
%(azimuthally weighted to take into account the detector acceptance) 
of the center of the azimuthal window observed to contain
an exotic photon or charge excess region. The regions are observed to be distributed
throughout the full azimuth, indicating that the exotic 
regions are observed throughout the entire acceptance regions of the detectors.

The second check concerns the investigation of possible time-specific detector readout
malfunctions. For each event found to have an exotic region, the neutral 
particle fraction ($f$) was calculated in the immediately preceding and immediately succeeding 
events in the same ($\Delta\eta$,$\Delta\phi$) region. The $f$ distributions of these events associated
to corresponding exotic events with $f_{max}$ values exceeding the $3\sigma$  cut 
or $f_{min}$ values 
less than the $3\sigma$ cut are shown in Fig.~\ref{fig6} (center). It is observed that the associated  
$f$ distributions have their peak around 1/3 and exhibit  behaviour similar to 
that of generic events, which in turn suggests normal detector operation.

The third check investigated the operation of the SPMD by examining if the
observed variation in the number of charged particles in an exotic region extended
to the pseudorapidity region 2.35~$ < \eta <$~2.9 of the SPMD that does not overlap 
with the PMD. 
Fig.~\ref{fig6} (bottom) compares the charged particle multiplicity distributions
in the non-overlapping part of the SPMD in the same azimuthal zone as the 
exotic region, and in the same event,
with the distributions in a randomly selected region of similar size, 
for photon excess as well as for charge excess regions. 
The good agreement of the
charged particle multiplicity distributions for regions near the exotic location and 
random regions suggests normal operation of the SPMD when the exotic regions are found.

In summary, photons and charged particles have been measured in a common $\eta$ - $\phi$ phase
space region in the WA98 experiment for 158 A GeV Pb+Pb collisions. The results were 
analyzed for the 15\% most 
central collisions to search for $\Delta\eta$-$\Delta\phi$ regions of photon or charged multiplicity 
excess using the sliding window method. The neutral particle
fraction ($f$ ) has been calculated in $\Delta\eta$-$\Delta\phi$ regions within 
2.9$ < \eta <$ 3.75 for different $\Delta\phi$ windows. It is found that regions 
with photon or charged multiplicity excess occur more frequently 
in the data than in fully mixed events or in VENUS+GEANT simulation events.
The percentage of events with photon-excess or charge-excess increases with decreasing
$\Delta\phi$ bin size, but varies little if at all as a function of centrality within the 
15\% most central collision event sample.
The exotic regions having large charged-neutral fluctuations are distributed throughout 
the azimuthal acceptance. Several investigations have been performed to rule out 
possible detector artifacts. The comparison with the analysis of mixed events provides a model
independent demonstration of non-statistical charged-neutral fluctuations.

%\begin{acknowlegement}

%\begin{sloppypar}
We wish to thank the CERN accelerator division for the excellent performance of the SPS
accelerator complex. We acknowledge with appreciation the effort of all engineers,
technicians, and support staff who have participated in the construction of this experiment.
This work was supported jointly by
the German BMBF and DFG,
the U.S. DOE,
the Swedish NFR and FRN,
the Dutch Stichting FOM,
the Polish MEiN under Contract No. 1P03B02230 and CERN/88/2006
The Grant Agency of the Czech Republic under contract No. 202/95/0217,
the Department of Atomic Energy,
the Department of Science and Technology,
the Council of Scientific and Industrial Research and
the University Grants
Commission of the Government of India,
the Indo-FRG Exchange Program,
the PPE division of CERN,
the Swiss National Fund,
the INTAS under Contract INTAS-97-0158,
ORISE,
Grant-in-Aid for Scientific Research
(Specially Promoted Research \& International Scientific Research)
of the Ministry of Education, Science and Culture,
the University of Tsukuba Special Research Projects, and
the JSPS Research Fellowships for Young Scientists.
ORNL is managed by UT-Battelle, LLC, for the U.S. DOE
under contract DE-AC05-00OR22725.
The MIT group has been supported by the U.S. Dept. of Energy under the
cooperative agreement DE-FC02- 94ER40818.
%\end{sloppypar}
%\end{acknowledgement}

\end{twocolumn}

\end{document}